\documentclass{article}
\usepackage{graphicx} 
\usepackage{pgfplots}
\pgfplotsset{compat=1.18}
\usepackage{tikz}
\usepackage{amsmath,amsfonts,amssymb}
\usepackage{subcaption}
\usepackage[margin=1in]{geometry}
\usepackage{parskip}
\usepackage{indentfirst}
\usepackage{authblk}
\usepackage{setspace}
\usepackage{todonotes}

\usepackage[sorting=none, citestyle=numeric-comp]{biblatex}
\usepackage{biblatex2bibitem}
\bibliography{References}

\usepackage{enumitem}

\title{\textbf{A Versatile Laboratory Approach to Reproduce and Analyze Internal Ocean Wave Dynamics}}
\begin{document}

\author[1]{Vohn Jacquez}
\author[1]{Zachary Phan}
\author[2]{Zachary Taebel}
\author[3]{Dylan Bruney}
\author[4]{Pierre-Yves Passaggia}
\author[5]{Alberto Scotti}
\affil[1]{University of North Carolina, Chapel Hill, NC, USA}
\affil[2]{Scripps Institute of Oceanography, San Diego, CA, USA}
\affil[3]{Wake Forest University, Winston-Salem, NC, USA}
\affil[4]{Univ. Orl\'eans, INSA CVL, PRISME, UPRES 4229, Orl\'eans, France}

\affil[5]{Arizona State University, Phoenix, AZ, USA}

\affil[ ]{Email address for correspondence: vbjacq8@unc.edu}

\date{\today}

\maketitle

\begin{abstract}
    Internal waves, or waves that propagate within a stratified fluid, may break and cause mixing. While each individual mixing event may be small, collectively, internal wave breaking  drive processes in the ocean that are critical to understanding the maritime climate and biosphere. In this paper we show how to set up  an experiment, suitable for an undergraduate-level lab, that illustrates a common generation and breaking mechanism in the ocean. In particular, we show how the process changes in response to a non dimensional parameter, the buoyancy Reynolds number, that can be easily varied. This parameter highlights the role of viscous vs. inertial/buoyancy forces.   We outline our methods of creating a linear stratification, injecting energy with a forced topography, and analyzing the resulting dynamics with Background Oriented Schlieren and energy spectra from a conductivity probe.  By altering our forcing to accommodate three values of the buoyancy Reynolds, three distinct internal wave regimes can be observed: no turbulence, slight turbulence, and extreme turbulence. Our methods aim to increase the accessibility to studying these internal waves in future experimental work, ocean modeling, and math and physics undergraduate learning.
\end{abstract}

\section{Introduction}
    Any child on the beach may notice how waves at the surface crash and displace shells and pebbles by the shore. However, surface waves are not the only type of waves found in the oceans. The underwater abyss is home to waves with amplitudes reaching hundreds of meters—these \textit{internal waves} are ubiquitous throughout the ocean interior, with notable hotspots  including the South China Sea, the Bay of Biscay, and along the Mid-Ocean ridges\supercite{mackinnon2013diapycnal}. Internal waves induce the vertical mixing of nutrients, heat, and dissolved gases through wave breaking\supercite{mackinnon2017climate}. This mixing has various significant effects at regional levels, including mitigating coral bleaching\ \supercite{wyatt2020heat}, upwelling nutrients into the photic zone\supercite{gaxiola2002internal}, and regulating melting rates of Arctic glaciers\supercite{washam2020tidal}. 
    
    Thus, studying how internal waves behave in our oceans is quite valuable. However, research cruises to study them in the ocean are costly and time-consuming. Furthermore, data collection often requires expensive arrays of many instruments that sample for long periods of time. Fortunately, one can recreate select dynamics of internal ocean waves in a laboratory using a relatively simple setup. This enables answering relevant questions of their generation, propagation, and induced mixing all within a wave tank. In addition, it visually exemplifies the peculiar propagating behavior of internal waves. Unlike other common types of waves, internal waves propagate energy at right angles relative to their phase speed.  
    
    In this paper, we describe how to set up an experimental apparatus that can be used to generate and observe the evolution of internal waves. The setup does not require exotic hardware. With a very modest investment of time and material, it is possible to achieve a qualitative view of the process. A little more effort yields quantitative measurements, which can be used, in addition to study internal waves, to illustrate concepts such as power spectra, phase and group velocity of waves, etc.  
    The paper is organized as follows: We begin with a description of generating a continuous stratification to model the abyssal ocean density profile. Next, we discuss how to replicate internal wave generation in the ocean, specifically internal waves born from tidal flow past rough seafloor topography. This excites internal waves at the tidal frequency, known as \textit{internal tides}\supercite{sutherland2010internal}. We recreate internal tide formation by oscillating an idealized model ocean ridge through our stratified fluid. Finally, we discuss data collection for these experiments, using both localized measurements of conductivity and optical methods to obtain global information on the wave field. 

    A benefit of this simple approach is its versatility in exploring different regimes of internal wave dynamics. It is suitable for analyzing both simple linear wave theory, weakly nonlinear wave-wave interactions, and turbulent wave-driven mixing. To aid in categorizing these regimes, we introduce a novel and unique way of estimating the buoyancy Reynolds number $Re_b$, based on linear wave theory and defined in terms of forcing parameters. We demonstrate the utility of our Reynolds number and our overall methodology by presenting results from three experiments that span roughly two orders of magnitude $Re_b$. We find that all configurations produce linear internal waves, but nonlinearities and resulting general turbulence increases as $Re_b$ increases. Thus, we are confident that our setup can explore the relevant range of regimes desired for different aspects of the internal wave problem, and that our approximation of $Re_b$ can aid in classifying them.
    
    Given that we are able to produce internal waves, access different regimes, and analyze dynamics, our experimental procedure is versatile, cost-effective, and simple. The nonlinear interactions and turbulence produced by our setup offer valuable topics for students hoping to do research in the fields of applied physics and mathematics, as these internal waves are dispersive and carry unique properties compared to light and sound waves which are traditionally covered in the classroom. Furthermore, the relevance of internal waves to our oceans and climates justify studying them from the lens of environmental and marine sciences in order to develop our climate and ocean models.
    
\section{Theory}
\subsection{Characterizing the Abyssal Ocean Interior}
Before revealing how to produce and study internal waves, we introduce several relevant parameters that characterize the domain in which they live and the resulting wave dynamics. We consider internal waves that exist within a continuously stratified fluid, such that the density $\rho = \rho(\mathbf{x},t)$. For laboratory studies, it is often convenient to treat the system as two-dimensional, with a horizontal direction aligned with the forcing\footnote{While internal waves in the ocean are three-dimensional, they are usually horizontally isotropic, meaning their horizontal direction does not affect their dynamics. Thus, their relevant behavior can be captured in 2D.}, so $\mathbf{x} = (x,z)$. 

Another useful simplification is to decompose $\rho$ using the Boussinesq approximation\supercite{cushman2011introduction} as $\rho(\mathbf{x},t) = \rho_0 + \bar{\rho}(z) + \rho'(\mathbf{x},t)$. Here $\rho_0$ is a constant reference density (often a mean density), $\bar{\rho}(z)$ is the vertical background stratification, and $\rho'(\mathbf{x},t)$ represents perturbations to the background state caused by the presence of waves. By definition, $\rho_0 + \bar{\rho} \gg \rho'$. We can then quantify the strength of the stratification by the Brunt-V\"ais\"al\"a (BV) frequency
\begin{equation}
    N = \sqrt{-\frac{g}{\rho_0}\frac{d\bar{\rho}}{dz}},
\end{equation}
which is the frequency a fluid parcel would oscillate, if displaced from its neutral density surface\supercite{sutherland2010internal}. The restoring force can also be expressed through a \textit{buoyancy perturbation} (often abbreviated as just ``buoyancy") $b = -g\rho'/\rho_0$, which is a measure of (minus) the reduced gravity, similar to the Archimedean concept of a buoyant force\supercite{serway2000physics}. 

As shown in Appendix A, under these simplifications, a traveling wave is a solution to the equations of motion provided its frequency $\omega$ and wavevector $\mathbf{p} = (k,m)$ obey the dispersion relation 
\begin{equation}
    \omega^2 = N^2\frac{k^2}{k^2 + m^2}.
    \label{eq:IWDispRel}
\end{equation}

The dispersion relation \eqref{eq:IWDispRel} has some unique properties, which are more visible by defining the wave vector in polar coordinates, using $k = p\cos(\theta)$ and $m = p\sin(\theta)$, with $p = |\mathbf{p}|$ and $\theta$ as the angle between $\mathbf{p}$ and the horizontal. Under this substitution, \eqref{eq:IWDispRel} becomes 
\begin{equation}
    \omega^2 = N^2\cos^2(\theta).
    \label{eq:DispRelTheta}
\end{equation}
Expressing the dispersion relation in this form presents three important takeaways. First, $\omega\leq N$, which sets an upper limit on wave frequencies. Second, a specific frequency corresponds to a specific angle of propagation. Lastly, $\omega$ is not related to the magnitude of $\mathbf{p}$, only its direction\footnote{ Contrast this to the dispersion relationship for sound or light waves propagating in an isotropic medium: in the latter case, the frequency depends on the magnitude of $\mathbf{p}$, rather than the angle $\theta$.}.  Thus, \eqref{eq:IWDispRel} has an infinite number of solutions $(k,m)$ which all have the same ratio $m/k$, since $\theta = \mathrm{arctan}(m/k)$. Consequently, applying a forcing at a single frequency, such as a tidal frequency, can excite many different spatial modes. The superposition of these modes propagates at the group velocity $\mathbf{c}_g = (\partial \omega/\partial k,\partial\omega/\partial m)$. From \eqref{eq:IWDispRel}, it can be shown that $\mathbf{c}_g$ is orthogonal to $\mathbf{p}$, so the superposition of waves propagates at an angle $\phi$ perpendicular to $\theta$. Therefore, the direction of the wave group is set by 
\begin{equation}
    \omega^2 = N^2\sin^2(\phi),
    \label{eq:IWDispRelAlpha}
\end{equation}
and the group propagates with a slope $\alpha$ given by
\begin{equation*}
    \alpha^2 = \frac{\omega^2}{N^2 - \omega^2}.
\end{equation*}

\subsection{Parameters of Nonlinearity}
Our focus is on the internal tide: internal waves formed when tidal flow runs over rough ocean topography. This process fuels roughly half the internal wave energy in the oceans \supercite{whalen2020internal}. If we are to recreate natural internal tides in a laboratory system, we must construct our forcing with the established parameters of nonlinearity in mind. 

To mathematically describe internal tide generation, we apply the linear wave theory of the prior section to a domain with a seafloor ridge of variable height $h(x)$ (with maximum height $h_0$), and a background oscillatory tidal flow of amplitude $A$, frequency $\omega_0$, and peak velocity $U_0=\omega_0 A$. This formulation dates back to Bell (1975)\supercite{bell1975lee}, and has been expanded by several other studies\supercite{llewellyn2002conversion,balmforth2002tidal,petrelis2006tidal}. The seafloor ridge produces a flow that is characterized by two key parameters. The first is the excursion number, defined as

    \begin{equation}
        Ex = \frac{A}{l_0}
    \end{equation}

where $l_0$ is a characteristic horizontal length scale of the topography, which we consider to be half its length. For small values ($Ex\ll 1$), the flow response to the tidal forcing is almost exclusively the internal tide. As $Ex$ increases, some of the tidal energy is converted to harmonics of the internal tide\supercite{garrett2007internal}, while large values of $Ex$ feature secondary dynamics such as lee waves and overturning\supercite{jalali2014tidal}. Fortunately, $Ex\ll 1$ for most of the large ocean ridges that generate internal tides\supercite{garrett2007internal}.

The second parameter of nonlinearity is the slope criticality, originally defined as 
\begin{equation*}
        \bar{c} = \frac{h_0}{l_0\alpha}, 
\end{equation*}
which is a ratio of an approximate ridge slope $h_0/l_0$ to the propagation angle of the wave group $\alpha$. For more complex topography, a better definition is to instead use the maximum slope of the ridge, such that 
\begin{equation}
    c = \frac{\max\left(dh/dx \right)}{\alpha}.
\end{equation}
The criticality compares the trajectory of fluid parcels as they are swept along the ridge against the natural trajectory of the waves at frequency $\omega_0$. Regimes where $c<1, c=1,$ and $c>1$ are referred to as subcritical, critical, and supercritical, respectively. Subcritical topographies correspond to the most linear tidal conversion, and are most typical of large ocean ridges\supercite{garrett2007internal}. Increasing $c$ tends to increase the number of excited spatial harmonics, leading to elevated shear and turbulence near the topography\supercite{sarkar2017topographic}. 

However, one more parameter is needed. We have not mentioned viscosity because the theory is derived for inviscid flow, but experiments occur at scales where viscous effects cannot be entirely neglected. Thus, it is beneficial to quantify these impacts through a Reynolds number, $Re$, a ratio of the inertial to viscous effects in the flow.  When $Re \ll 1$, viscosity smothers out nonlinearities, while when $Re \gg 1$, the flow can become turbulent.  For the internal tide problem at hand, the simplest definition uses the topographic height $h_0$ as the length scale and sets maximum tidal velocity $A\omega _0$ as the velocity scale. With a kinematic viscosity $\nu$ we arrive at
\begin{equation}
        Re = \frac{UL}{\nu} = \frac{A\omega h_0}{\nu} = \frac{Ex l_0\omega h_0}{\nu}
    \end{equation}
This definition suggests that viscous effects are mitigated through a large tidal amplitude, a large frequency, and/or a tall ridge, which is supported by previous internal wave studies \supercite{jalali2014tidal}. This 'forcing' Reynolds number presents one way to quantify the effects of viscosity on system energy dissipation and mixing (increased viscosity smothers out any nonlinear interactions and potential for mixing). It also takes into account the previously defined excursion number.

\subsection{Deriving a novel buoyancy Reynolds number}

The Reynolds number discussed in Section 2.2 can be used to infer turbulence near the topography, but internal wave driven mixing often occurs due to wave breaking in the far field. Furthermore, in a stratified fluid, turbulence is affected by both viscosity and stratification. We therefore want a way to estimate the turbulence caused by waves in the far field while accounting for the effects of stratification. 

To begin, we introduce two important length scales that help assess the influence of stratification and viscosity. The first is the Ozmidov length scale $L_O = \sqrt{\epsilon/N^3}$, which characterizes the largest turbulent eddies that are not strongly suppressed by buoyancy. The second scale is the Kolmogorov scale $L_K = (\nu^3/\epsilon)^{1/4}$, which characterizes the smallest eddies that persist before being dissipated by viscosity\footnote{A more detailed explanation of the Ozmidov and Kolmogorov length scales is provided in Appendix B.}. By dividing these two scales and raising the result to the power of $3/4$, we obtain a dimensionless number that compares the turbulent energy dissipation rate $\epsilon$ to the strength of viscosity and stratification. This number is known as the buoyancy Reynolds number $Re_b$, defined as
\begin{equation}
    Re_b = \left(\frac{L_O}{L_K}\right)^{\frac{3}{4}} = \frac{\epsilon}{\nu N^2}.
\end{equation} 
Thus, large values of $Re_b$ reflect turbulent flow that overcomes the restraints of buoyancy and viscosity, while small values of $Re_b$ mean that one (or both) of the mitigating forces significantly dampen turbulence. 

One challenge in quantifying $Re_b$ is the difficulty in estimating $\epsilon$ experimentally. However, recall our interest is in turbulence caused by internal wave breaking. If we assume that all energy pumped into waves by the tide eventually ends up fueling turbulence, then we can estimate $\epsilon$ based on the energy input to the internal tide. We start with an internal tide energy flux away from the site of generation, $F_\text{Linear}$, modeled by Llewellyn Smith and Young\supercite{llewellyn2002conversion} with slight alterations:
\begin{equation}
     F_{\text{Linear}} = \frac{1}{2\pi}\rho_0  \omega^{-1} \sqrt{(N^2-\omega ^2)(\omega^2-f^2)}U_{0}^2 \alpha \pi H^{-1}\sum_{n=1}^\infty k_n\left|\frac{\tilde{h}(k_n)}{h_0}\right|^2    
\end{equation}
where 
\begin{align*}
    \alpha &= \sqrt{\frac{N^2-\omega^2}{\omega^2}} \ (\text{inverse of beam slope}), & k_n&=\frac{\alpha\pi n}{H}.
\end{align*}
Here, $k_n$ is the discretized horizontal wavenumber corresponding to mode $n$ through the dispersion relation, and $\tilde{h}(k)/h_0$ is the Fourier transform of our topographic profile $h(x)$. We neglect the Coriolis frequency $f$ because its effects are negligible in our experiments, condense terms for notation, and simplify to arrive at 
\begin{equation}
    F_{\text{Linear}} = \frac{1}{2}\rho_0N\sqrt{1-\frac{\omega^2}{N^2}}U_{0}^2 h_0^2\alpha G(h),
\end{equation}
with 
\begin{equation}
    G(h) = H^{-1}\sum_{n=1}^\infty k_n\left|\frac{\tilde{h}(k_n)}{h_0}\right|^2.
\end{equation}

To convert from the dimensions of $F_{Linear}$ to $\epsilon$, or energy dissipation, we must divide by a mass and multiply by a length. We divide $F_\mathrm{Linear}$ by$\rho_0LH$ where $L$ is the domain length. We also expand $U_0 = A\omega$ to obtain
\begin{equation}
    \epsilon = \frac{1}{2} \frac{1}{LH}N\sqrt{1-\frac{\omega^2}{N^2}}A^2\omega^2 h_0^2\alpha G(h).
\end{equation}
Finally, to get a buoyancy Reynolds number $Re_b$, we divide by $N^2\nu$ to get
\begin{equation}
    Re_b = \frac{1}{2N\nu}\frac{1}{LH}\sqrt{1-\frac{\omega^2}{N^2}}A^2\omega^2 h_0^2\alpha G(h).
\end{equation}
This Reynolds number can be expressed in terms of the excursion number $Ex$ by multiplying and dividing by $l_0^3$:
\begin{equation}
    Re_b = \frac{1}{2N\nu}\frac{l_0 h_0}{LH}\sqrt{1-\frac{\omega^2}{N^2}}Ex^2\omega^2 \frac{h_0}{l_0}\alpha G(h).
\end{equation}
Recall that $h_0/l_0$ is an approximate measure of the topographic slope. Because $\alpha$ is the inverse of the beam slope, the product $\alpha h_0/l_0$ is the slope criticality $\bar{c}$.
\begin{equation}
    Re_b = \frac{1}{2N\nu}\frac{l_0 h_0}{LH}\sqrt{1-\frac{\omega^2}{N^2}}Ex^2\bar{c} \omega^2 G(h).
\end{equation}
If needed, one can use the dispersion relation and a trigonometric identity to further simplify:
\begin{equation*}
    Re_b = \frac{1}{2N\nu}\frac{l_0 h_0}{LH}\cos({\phi})Ex^2\bar{c}\omega^2G(h).
\end{equation*}
Thus, we are able to create a holistic estimate of turbulence with a buoyancy Reynolds number that incorporates linear wave theory as well as existing parameters of nonlinearity. Furthermore, it is defined entirely in terms of known parameters, enabling estimation before performing experiments to aid in the setup choices. It should be mentioned that this approximation of $Re_b$ is likely smaller than the true buoyancy Reynolds number, since energy may fuel turbulence from sources other than linear waves. Furthermore, our approximation is a global estimate, while $Re_b$ may locally be much larger. Still, we will show that even as a lower bound, this approximation is a powerful aid in predicting and characterizing the flow response. 

\section{Methods}

\subsection{Materials Used}
Creating internal waves in a laboratory first requires a box to act as a wave tank to model ocean behavior, which should be made of a transparent material to visualize the waves. We chose acrylic as the material due to the low price point and safety over glass, with tank dimensions of 203cm long x 8cm wide x 46cm tall. The exact size precision is not particularly important. Rather, keeping the tank relatively thin and having a large ratio between length and width, restricting flow within the desired two dimensions is more important. We used three other holding tanks with non-specific dimensions: one for housing saltwater, one for housing freshwater, and one to mix the two to create a stratified fluid. We then utilized peristaltic pumps and a mixing device to move and blend fresh water and brine in the appropriate ratio to fill the experimental tank with a stratified fluid.  A controllable motor system is necessary to move the topography back and forth with tunable frequencies and amplitudes. Our experiments use a Parker Automation stepper motor, but a simple electric motor with a wheel and pin acting as a crankshaft can do the job. If one is interested in a simple qualitative view of internal wave dynamics, they can place a strong light source in front of the wavetank to shadowgraph the waves onto a board (our experiments use a heat lamp). The light source must send parallel rays, which can be achieved by placing the source at the focal point of a concave mirror or reflective surface. However, for a more quantitative view, it is useful to procure a conductivity probe and high-resolution cameras to be used for BOS later on. 

\subsection{Setup and Procedure}
To emulate the interior ocean and create a domain for the waves to propagate, the tank must first be filled with a stratified fluid. We recommend saltwater, as rock salt provides an inexpensive, accessible, and safe stratifying agent. The chosen salt should not contain anti-caking agents, as these can reduce the transparency required for optical analysis. 

\begin{figure}
    \centering
    \includegraphics[]{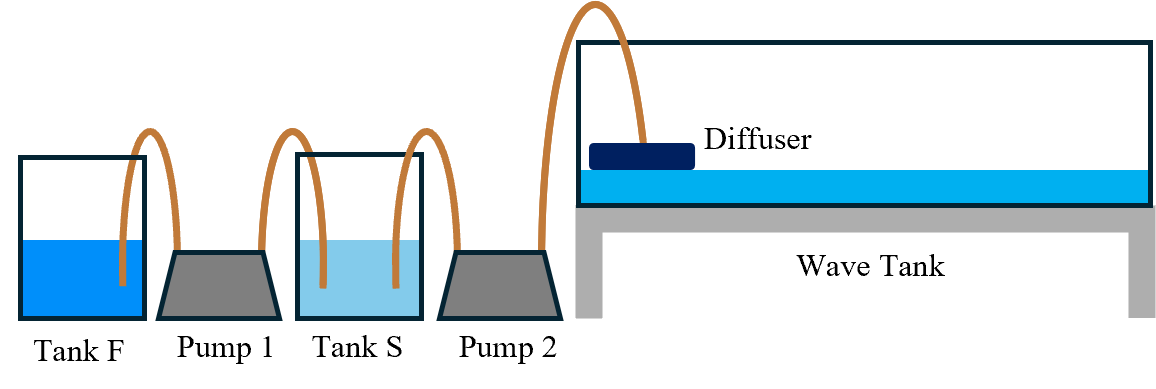}
    \caption{Schematic of the `two-bucket' method. Pump 1 pumps freshwater from tank F into tank S (which has been previously filled with saltwater) at rate $R_1$. A bilge pump (not shown in the schematic) in Tank S mixes the tank vigorously.  Pump 2 simultaneously moves the mixed solution through the diffuser and into the tank at rate $R_2=2R_1$.}
    \label{fig:twobucketschematic}
\end{figure}

\begin{figure}
    \centering
    \includegraphics[width=6in]{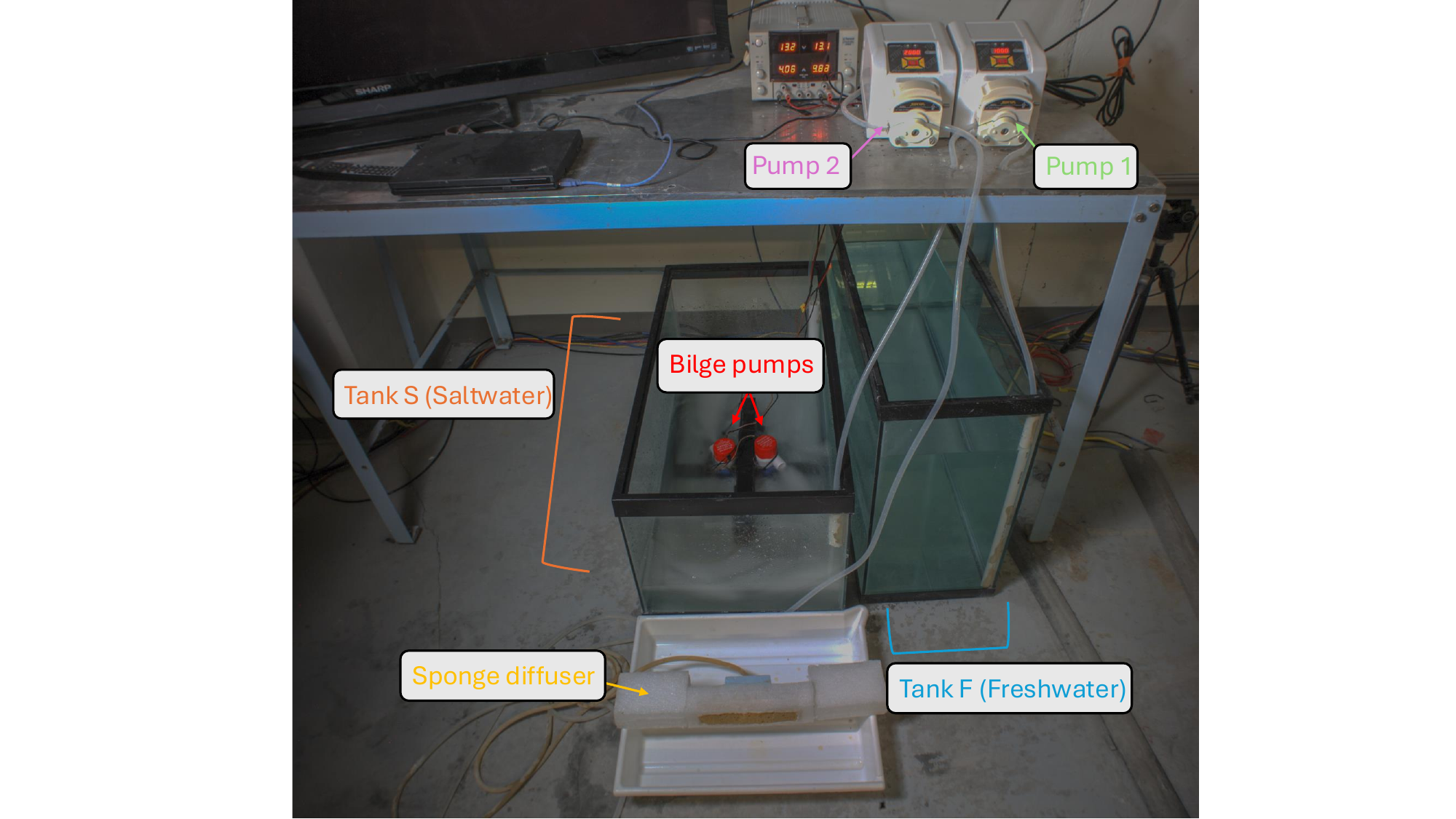}
    \caption{Experimental setup utilizing the `two-bucket' method. Pumps 1 and 2 (BT-300EA JIHPUMP) move water through Tanks S and F as bilge pumps (Maxzone Bilge Pump 1100GPH Mod. A01-111001) mix water in Tank S. Diffuser is made of sponge and floats on the water level as water is pumped into tank. }
    \label{fig:twobucketexperiment}
\end{figure}

After creating saltwater in the off-hand storage tank, the solution is transferred to a mixing tank. 
We then use this solution to create a linear stratification through the two-bucket method\supercite{oster1963density}. In this method, freshwater is slowly pumped into the mixing tank to be mixed with saltwater (which we accomplish by adding circulation with bilge pumps). The resultant mixture is simultaneously pumped  into the wave tank. This process is shown schematically in Fig. \ref{fig:twobucketschematic}, and a photo of our setup is shown in Fig. \ref{fig:twobucketexperiment}. As shown in the figures, we disperse the water added to the wave tank by running it through a sponge, which floats at the water surface. This avoids any unintentional mixing caused by water entering the tank. Another important step that may be considered is having some sort of brace on top of the wave tank as fluid enters, as students may observe that due to hydrostatic pressure, the center of the wave tank will bulge outwards.

Thus, the density at the bottom coincides with the density of the water in the saltwater tank at the beginning of the filling process. Since the mixing tank becomes less salty over time as more freshwater is pumped into the mixing tank the experimental tank fills with progressively lighter water. By setting the flow rates of pump 1 and pump 2 at a 1:2 ratio, the resulting density profile is linear with depth. A linear stratification is desirable because it features a constant $N$, and hence, a constant wave angle $\theta$ , which makes data analysis simpler.  To confirm the linear stratification and calculate $N$, we take discrete measurements with the conductivity probe and stepper motor throughout the depth of the tank on one end to minimize disturbances. We then use a conductivity-to-density spline fit calibrated with saline solutions of varying density to determine the stratification.

Emulating wave generation by tidal flow past a seafloor requires oscillating fluid past a stationary topography, which requires moving the walls of the wave tank back and forth. However, doing so involves an expensive piston-powered paddle machine strong enough to displace the water along the entire vertical edge, which is not practical nor affordable for many undergraduate labs. Fortunately, one can change the reference frame and oscillate the topography instead while keeping the fluid stationary, remaining dynamically equivalent while cutting complexity and costs, provided that the relative difference in buoyancy between top and bottom does not exceed 20\%.  The topography can be constructed with any nonporous material, such as Styrofoam. \supercite{echeverri2009low,spiegel1960boussinesq}.  For our topography, we used an idealized ridge defined by a hyperbolic secant squared:

    \begin{equation}
    h(x) = h_0\mathrm{sech}^2\left(\frac{x}{h_0}\right)-h_0\mathrm{sech}^2\left(\frac{l_0}{h_0}\right).
\end{equation}
Note that $h(x)=0$ when $x=\pm l_0$ where $l_0$ is the half-length of the topography. Defining the topography this way simplifies the calculation of the Fourier transformed term within our buoyancy Reynolds number $Re_b$.

We aim to demonstrate that our setup may accommodate unique internal wave dynamics by sweeping this buoyancy Reynolds number across three orders of magnitude: small $(Re_b = 0.0076)$, medium $(Re_b =0.0752 )$, and large $(Re_b  = 0.7256)$. These values are achieved by using three ridges with unique lengths, heights, and oscillation amplitudes, but are forced at a fixed excursion number and frequency. In the configurations for $Re =0.0076$ and $Re=0.0752$, we use one stratification with BV frequency $N = 1.3941 \ \text{rad} \ \text{s}^{-1}$ . We performed the experiment at $Re_b = 0.7256$ separately, with a similar BV frequency $N = 1.4516 \ \text{rad} \ \text{s}^{-1}$. Despite this discrepancy in stratification, the more important ratio $\frac{\omega}{N}=\sin\theta$ remains roughly constant. The details for all system parameters in each experiment are shown in Table I.
\begin{table}[h]
    \centering
     \begin{tabular}{|c|c|c|c|c|c|c|}
        \hline
            Configuration & $Re_b$ & $N$ ($\text{rad} \ \text{s}^-1$) & $\omega_0$ $(\text{rad} \ \text{s}^{-1})$& $Ex$ & $h_0$ (m) & $l$ (m) \\
            \hline
            1 & 0.0076 & 1.39414 & 1.0472 & 0.1562 & 0.0314 & 0.0858 \\
            \hline
            2 & 0.0752 & 1.39414 & 1.0472 & 0.1561 & 0.0558 & 0.1525 \\
            \hline 
            3 & 0.7256 & 1.45156 & 1.0472 & 0.1560 & 0.0992 & 0.2712  \\
            \hline
          
        \end{tabular}
\caption{Parameters used for all three experimental configurations of increasing $Re_b$ by changing oscillation spatial scales. Note that the excursion number and slope criticality are both constant for all experiments.}
\end{table}

\begin{figure}
    \centering
    \includegraphics[width=\linewidth]{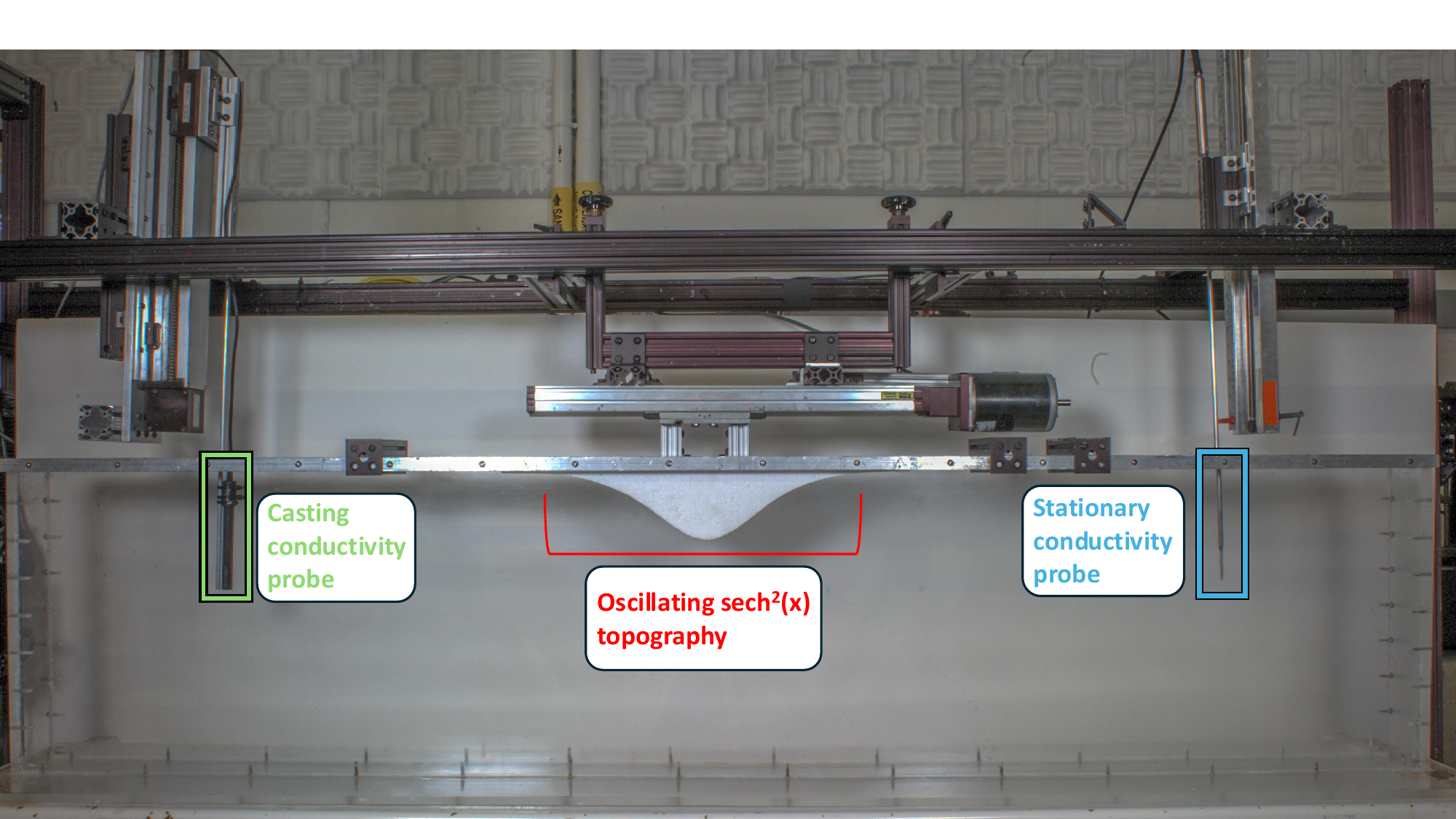}
    \caption{Wavetank apparatus for internal wave generation and measurement. Casting conductivity probe (left) yields a pre-experiment BV frequency. Waves are formed by oscillating the topography (center) with a stepper motor. Stationary conductivity probe (right) identifies energetic frequencies.}
    \label{fig: wavetank}
\end{figure}

\subsection{Data Collection}
Creating internal waves is only half the battle, because the waves themselves are invisible to the naked eye. Therefore, confirming their existence and analyzing their behavior requires additional pieces of equipment. For a purely qualitative approach we recommend a shadowgraph, where strong, parallel light shines through the tank onto a board on the other side. As internal waves propagate, they perturb the density field, causing light to refract and cast shadow patterns on the board. This low-cost technique (we use a heat lamp  and whiteboard) allows for a qualitative visualization of the wave field across the entire tank, but requires the minimization of ambient light. A live video demonstration is shown in Supplement Number 1.

For a more quantitative view, the simplest approach to data collection is to use localized density measurements obtained from a conductivity probe. The internal waves propagate from the topography at angles determined by the dispersion relation \eqref{eq:IWDispRelAlpha}, making their location easy to predict. By placing the probe within the trajectory of the internal waves, measuring the wave amplitude is achievable. Students may find it interesting to brush up on their trigonometric skills to determine where to place the probe as well! 

A limitation of such localized measurements is that they only provide information in time, but we often also desire information in space. This requires global data about the wave field, and doing so in the ocean necessitates an array of closely spaced, localized measurements along moorings. On the other hand, obtaining global information experimentally is far easier and non-intrusive because of optical methods that capitalize on the relationship between a fluid's index of refraction to its density gradient. A more quantitative method for retrieving the full density field is known as Background Oriented Schlieren (BOS), or Synthetic Schlieren. In BOS, cameras look through the tank at a stationary image on the other side. As internal waves propagate, the perturbations to the index of refraction distort the image that the cameras observe. By comparing a distorted image to a reference image, we can recover the density gradient through post-processing\supercite{passaggia2020estimating}. One can then numerically solve for the full density and velocity fields via linear wave theory.

Although BOS may be done with one camera, our setup uses three synchronized cameras to capture pictures of the entire tank. Behind the tank, the image typically used for BOS is a dot pattern on backlit paper. Here instead we use a computer projector to cast an image of a random dot pattern onto a blank background. This is cost-effective and convenient to set up, because the projector allows uniform illumination of the dot pattern. Furthermore, the distance between the projected image and the wavetank is easily increased by moving the whiteboard, which can improve BOS resolution \supercite{sutherland2014internal}.

\section{Results and Discussion}
    
To confirm our approach to internal tide generation, we present results from both BOS and localized conductivity measurements for our three experiments. We begin by demonstrating our creation of linear internal waves in Configuration 1 ($Re_b = 0.0076$). In the BOS, (Fig. 4), we observe clear and consistent diagonals in the density gradient, consistent with a wave beam as discussed in Section 2.1. At our forcing and BV frequency, the beam should propagate at an angle $\phi = 48.689^{\circ}$ below the horizontal, based on the dispersion relation. This trajectory is shown in Fig. 4 as superimposed arrows, which are roughly parallel to the observed diagonals. Given these agreements and the visual clarity of the beam, we are confident that this setup produces internal waves, specifically in the linear regime that is characterized by a small Reynolds number. 

\begin{figure}
    \centering
    \includegraphics[]{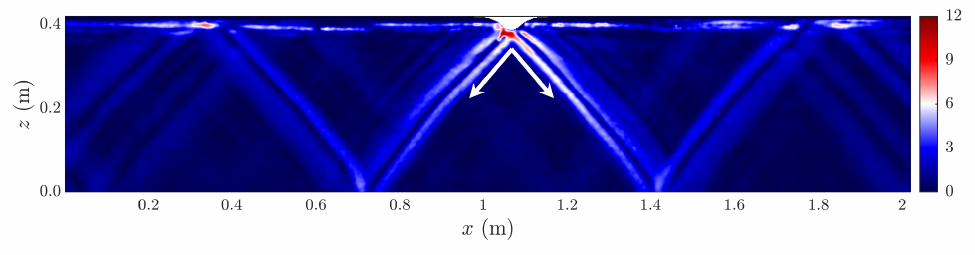}
    \caption{Magnitude of buoyancy gradient squared $|\nabla b|$ (in s$^{-2}$) obtained from BOS for Configuration 1, $Re_b = 0.0076$. The superimposed vector represents the expected angle $\phi = 48.689^{\circ}$ from the dispersion relation. Data taken from 452 seconds after the onset of forcing and smoothed using splines\supercite{garcia2010robust,gargett1981composite}.}
    \label{fig:enter-label}
\end{figure}

Our setup at a low $Re_b$ can produce internal waves that are useful for comparison with simple linear theory, but ocean dynamics are not so simple. To show that our setup can also achieve the nonlinear regime characteristic of ocean dynamics, we compare the prior results to those at larger $Re_b$ by plotting the buoyancy gradient magnitude for all experiments in Fig. 5. 

Figure 5(b) shows $|\nabla b|^2$ for $Re_b = 0.0752$. In this plot we still observe the predicted wave beam diagonal pattern, however, it is far less coherent. Inside the overarching diagonal there are several different slopes and small perturbations within the buoyancy gradient. At the next order of magnitude ($Re_b = 0.7256$), we are met with no overarching shape to be discerned in the BOS. Rather, the localized wave profile has been replaced by global turbulence. Note that here and thereafter when we say turbulence we mean wave turbulence, i.e. the weakly nonlinear process that generates waves at frequencies different than the forcing frequency.  

We supplement our analysis of these dynamics by considering the time evolution of each system. A video of this evolution is included within the Supplementary Material. In the low $Re_b$ regime, we see the wave beam propagate at a single angle for the full experimental duration. At medium $Re_b$, the buoyancy gradient initially displays a diagonal pattern at a single angle (similar to the low $Re_b$ dynamics), but then evolves into multiple angles with increased turbulence both near and away from the topography. This supports the idea that at larger $Re_b$, nonlinear effects are more intense. If we go one step further to the last configuration, the highest $Re_b$ system shows near-field turbulence from the start. The onset of forcing displays beam wave patterns, similar to the other experiments, but these patterns are quickly obfuscated by wave breaking and other nonlinearities. With time, the whole field becomes disordered. Thus, BOS analysis of our experiments demonstrates that we can escape the calm linear regime with a change in $Re_b$.

\begin{figure}[h]
    \centering
    \includegraphics[]{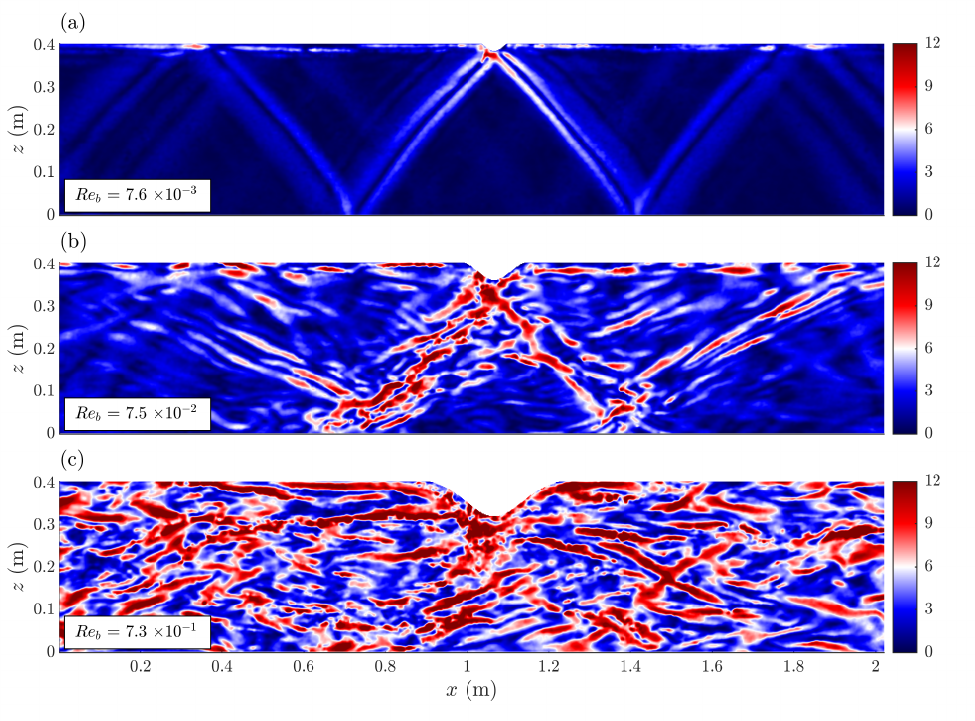}
   \caption{Magnitude of buoyancy gradient $|\nabla b|$ (in s$^{-2}$) obtained from BOS for (a) $Re_b = 0.0076$. (b) $Re_b = 0.0752$, and (c) $Re_b = 0.7256$. Topographic shape is superimposed. Data taken from 1835 s after the onset of forcing, and smoothed using splines\supercite{garcia2010robust,gargett1981composite}.}
    \label{fig:BOS3Panel}
\end{figure}

One point worth addressing is whether our approximation for $Re_b$ is still valid in the turbulent regimes at medium and large $Re_b$ as our derivation was based on linear wave theory. To determine whether the waves in our system are consistent with linear wave theory, we compare our BOS results against the dispersion relation. We first take Fourier transforms of the BOS data in space and time to obtain a 3D spectrum. We then average the spectrum over thin wedges of the angle $\theta$ to obtain a spectrum as a function of frequency and angle. This can be directly compared against the dispersion curve and is shown in Supplement Number 2. For all three experiments, we see that energy within the internal wave-permitting frequency range $(\omega < N$) preferentially lies along the dispersion relation, even at large $Re_b$. Therefore, even at large $Re_b$, our model of turbulence forced by linear waves serves as a viable representation of the system.

We now turn our attention to the energy spectrum of the density perturbations collected by our stationary conductivity probe (Figure 6), which determines how energy is partitioned into specific frequencies. The high sample rate of the probe allows us to investigate the frequencies below and above $N$. Thus we can examine how $Re_b$ is related to both the energy within the wave-permitting regime, and at turbulent frequencies which waves cannot access.

    \begin{figure}[h]
    \centering
    \hspace*{-1.8cm}
    \includegraphics[width=5in]{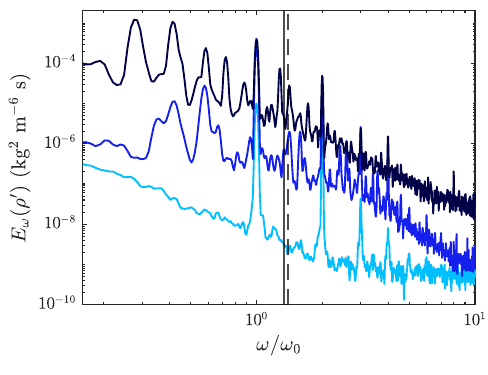}
    \caption{Energy spectrum of density perturbation from the stationary conductivity probe for (light blue) $Re_b = 0.0076$, (medium blue) $Re_b = 0.0752$, and (dark blue) $Re_b = 0.7256$. Spectra are formed using Welch's method with 8 segments of 50\% overlap, and then smoothed with a Gaussian smoothing function. Spectral values above 10 Hz are excluded due to excessive noise. Data taken over the full 30 minutes of forcing.}
    \label{fig:enter-label}
\end{figure}

For $\omega < N$, each setup differs in magnitude and quantity of discrete energy peaks. For our lowest $Re_b$ (light blue curve), the only energy peak occurs at the forcing frequency $\omega_0$. This agrees with there being only a single beam angle in the BOS results. This changes for the medium $Re_b$ (medium blue curve). In addition to the higher total energy, there are also two new peaks below $\omega_0$. This is attributed to nonlinear wave-wave interactions in which the internal tide births two `daughter' waves at lower frequencies\supercite{dauxois2018instabilities}. These frequencies correspond to the multiple wave beam slopes that are present in the BOS. At the highest $Re_b$ (dark blue curve), there is another increase in total energy. Additionally, there are five peaks below $\omega_0$ which follow the disordered activity in the BOS. Thus, our spectral results suggest that increasing $Re_b$ allows for more nonlinearity and wave-wave interactions. 
    
    For $\omega > N$, the distinction between $Re_b$ regimes lies in the magnitude of energy present. Starting at the highest $Re_b$, we see an elevated floor that depicts energy contained in high-frequency turbulence. This floor lowers as we look to our middle $Re_b$, which is initially comparable slightly past $N$ but quickly drops off at higher frequencies. If we look at our lowest $Re_b$, there effectively is no energy beyond $N$, except for several isolated peaks at harmonics of $\omega_0$. These harmonics cannot be linear internal waves since they occur at frequencies greater than $N$. These frequencies likely represent \textit{interfacial waves} at the base of a thin mixed layer produced near the forcing, which we observed due to the shallow depth of the probe. The appearance of these peaks is therefore not reflective of the far-field response, where no energy exists at frequencies above $N$.

    Thus, our BOS and energy spectra results show we can reliably produce a desired internal wave system that can vary based on the setup. Our variety of flow responses also demonstrates the efficacy of our approximation for $Re_b$ in characterizing specific internal wave dynamics. To recall, our derivation was based on linear wave theory. Because the theory and experimental results agree, we have shown a method of internal wave study that accounts for the whole system while being practical for lab experiments with constraints.

\section{Conclusions}
Understanding internal waves is crucial for understanding climatological and ecological processes due to the turbulent mixing they cause. However, because field measurements of wave-driven mixing in the deep sea are lengthy and expensive, we desire to study internal waves in a controlled  lab setting. Thus, we demonstrate a procedure to produce and analyze internal waves in a small lab tank. We constructed a simplified analogue of the ocean abyssal density profile using rock salt and the two-bucket method. We then recreate internal tide generation by oscillating an idealized topography through our tank in three experiments. In each, the flow response is characterized by our unique approximation for the buoyancy Reynolds number, obtained through linear wave theory. We observe and analyze our created internal waves with Background Oriented Schlieren for a global perspective and conductivity measurements for high-frequency energy spectra. Our results confirm the creation of linear internal waves, as well as confirm the effectiveness of our buoyancy Reynolds number in characterizing the extent of wave-driven mixing. Our experiments investigate $Re_b$ up to the order of $10^{-1}$, so future work may be done to explore $Re_b>1$. We anticipate these will likely demonstrate the same relationships between $Re_b$ and the presence of turbulence as our experiments do.

With this in mind, we suggest that these methods for creating and analyzing the internal waves can be easily replicated by undergraduate laboratories with diverse motivations: oceanography students curious to understand the processes which govern ocean mixing, fluid dynamics classes applying linear wave theory to a small system, or applied math and physics students interested in a real-world wave systems characterized by an unusual dispersion relationship. Researchers and students may find our procedure useful in creating internal waves as they can customize their setup to their desired dynamics. With the many `tunable knobs' in our Reynolds number, researching specific wave regimes may be done despite setup constraints. In combination with non-intrusive optical methods being our primary mode of analysis, we hope our work increases the accessibility of studying the internal waves that are crucially important to our Earth. 

\textcolor{blue}{}


\section{Author Declarations}
The authors have no conflicts of interest. 

\section*{Appendix A: Linear Wave Theory}

Under the definitions of $N$ and $b$ presented in Section 2, the Euler equations may be expressed as
\begin{subequations}
    \begin{align}
    \frac{\partial u}{\partial t} &= -\frac{1}{\rho_0}\frac{\partial P'}{\partial x}\\
    \frac{\partial w}{\partial t} &= -\frac{1}{\rho_0}\frac{\partial P'}{\partial z} + b\\
    \frac{\partial b}{\partial t} &= -wN^2\\
    \nabla \cdot \mathbf{u} &= 0.
    \end{align}
    \label{eq:EulerStrat}
\end{subequations}
To obtain these equations, we plug in the expanded definition of density $\rho(x,z,t) = \rho_0 + \bar{\rho}(z) + \rho'(x,z,t)$, and define the pressure as the sum of a hydrostatic background pressure and perturbations from this state: $P(x,z,t) = \bar{P}(z) + P'(x,z,t)$. The hydrostatic pressure obeys the relationship $\frac{d \bar{P}}{d z} =- (\rho_0 + \bar{\rho})$, which causes all background terms to cancel from the Euler equations, except for the continuity equation. 

We can dramatically condense the system of equations \eqref{eq:EulerStrat} by subtracting an $x$ derivative of (\ref{eq:EulerStrat}b) from a $z$ derivative of (\ref{eq:EulerStrat}a). Doing so returns
\begin{equation}
    \frac{\partial}{\partial t}\left(\frac{\partial u}{\partial z} - \frac{\partial w}{\partial x}\right) = -\frac{\partial b}{\partial x}.
    \label{eq:WaveSetup1}
\end{equation}
To reduce the number of variables, we introduce the \textit{streamfunction} of the flow $\psi$, which is related to the velocity through $u = -\frac{\partial \psi}{\partial z}$ and $w = \frac{\partial \psi}{\partial x}$. Through these definitions, the term inside the parentheses is equal to $-\nabla^2\psi$, such that \eqref{eq:WaveSetup1} may be reduced to 
\begin{equation}
    \frac{\partial}{\partial t}\nabla^2\psi = \frac{\partial b}{\partial x}.
    \label{eq:WaveSteup2}
\end{equation}
We can use the continuity equation to simplify further. By taking an $x$ derivative of (\ref{eq:EulerStrat}c), we obtain 
\begin{equation}
    \frac{\partial^2 b}{\partial t\partial x} = -\frac{\partial^2 \psi}{\partial x^2}N^2.
        \label{eq:WaveSetup3}
\end{equation}
We then take a $t$ derivative of \eqref{eq:WaveSteup2}, such that we can substitute the RHS of \eqref{eq:WaveSetup3} for $\frac{\partial^2 b}{\partial t\partial x}$. Doing so finally produces the internal wave equation
\begin{equation}
    \frac{\partial^2}{\partial t^2}\nabla^2\psi + N^2\frac{\partial^2 \psi}{\partial x^2} = 0.
    \label{eq:IWequationPSI}
\end{equation}
To seek out solutions, we plug in a simple traveling wave of the form $\psi(\mathbf{x},t) = \Psi e^{i(kx + mz - \omega t)} +c.c.$ to \eqref{eq:IWequationPSI}. By virtue of the simple derivative properties of exponentials, this returns
\begin{equation}
    \omega^2(k^2 + m^2)\psi - N^2k^2\psi = 0.
    \label{eq:IWcondition}
\end{equation}
From \eqref{eq:IWcondition}, we determine that simple travelling waves are indeed solutions to \eqref{eq:IWequationPSI}, provided the wavenumbers and frequencies obey 
\begin{equation*}
    \omega^2 = N^2\frac{k^2}{k^2 + m^2},
\end{equation*}
which is the dispersion relation \eqref{eq:IWDispRel} introduced in section 2. 

It is worth mentioning that the linearity of the system means that any of the original dependent variables from \eqref{eq:EulerStrat} have traveling wave solutions similar to those for $\psi$. For example, $u = -\frac{\partial \psi}{\partial z} = -im\psi$ and $w = \frac{\partial \psi}{\partial x} = ik\psi$. 

\section*{Appendix B: Ozmidov and Kolmogorov Length Scales}
Kolmogorov's formulation of turbulence envisioned a range of length scales within which the system is fully isotropic, that is, the spectrum depends only upon the magnitude of the wavevector and not its direction\supercite{kunze2019unified}. This range of scales is bounded (at the lower end) by the Kolmogorov scale, at which point the Reynolds number of the system equals 1. The Kolmogorov scale is defined as $L_K = \left(\nu^3/\epsilon\right)^{1/4}$. At scales smaller than this, viscous effects are more significant than inertial effects, and eddies systematically lose energy to viscosity. 

Stratification introduces an upper limit to the range of length scales that engage in isotropic turbulence. This occurs because buoyant forces will impede the vertical motions of eddies, but not horizontal motions (assuming horizontal isopycnals). This naturally introduces an anisotropy to the system, since horizontal motions can occur with greater freedom than vertical motions. However, at a certain point, an eddy may be small enough that the displacements caused by vertical motions induce a buoyant force that is too weak to actually alter the flow. To see this, note that in a continuously stratified fluid, a parcel of fluid that is displaced from its neutral density surface will experience an opposing buoyant force. This force increases in strength as the parcel is displaced further. One could then envision that for small enough displacements, the buoyant force will be too weak to smother vertical motions, and the eddies remain isotropic. We characterize this transition at the Ozmidov scale $L_O = \sqrt{\epsilon/N^3}$, when inertial and buoyant forces are of equal magnitude (for a derivation of $L_O$, see the appendix of Riley and Lindborg\supercite{riley2008stratified}). 

Provided $L_O > L_K$, there exists a range of scales that are small enough to be unaffected by stratification, but large enough to unaffected by viscosity. Within this range of scales, isotropic Kolmogorov turbulence can occur. This makes $Re_b$ a natural parameter for characterizing Kolmogorov turbulence, since $Re_b = \left(L_O/L_K\right)^{3/4}$, and thus values of $Re_b$ greater than 1 represent regimes in which isotropic turbulence is possible. On the other hand, if $Re_b < 1$, then an eddy would be dissipated by viscosity before it became small enough to be unaffected by stratification.

\end{document}